# Time delays and energy transport velocities in three dimensional ideal cloaking

Huanyang Chen,[1, 2, *] and C. T. Chan[2]

[1]*Department of Physics, Hong Kong University of Science and Technology, Clear Water Bay, Kowlong, Hong Kong, China*

[2]*Institute of Theoretical Physics, Shanghai Jiao Tong University, Shanghai 200240, China*

[*]*Corresponding author: moroshine@hotmail.com*

We obtained the energy transport velocity distribution for a three dimensional ideal cloak explicitly. Near the operation frequency, the energy transport velocity has rather peculiar distribution. The velocity along a line joining the origin of the cloak is a constant, while the velocity approaches zero at the inner boundary of the cloak. A ray pointing right into the origin of the cloak will experience abrupt changes of velocities when it impinges on the inner surface of the cloak. This peculiar distribution causes long time delays for beams passing through the ideal cloak within a geometric optics description.

There is growing interest for the invisibility cloaking both theoretically[1-14] and experimentally[15]. While invisibility cloaks are designed for one single operation frequency[13, 16, 17], partial invisibility cloaking for a wider bandwidth can be designed[17]. The perfect cloaking effect for both two dimensional (2D) and three dimensional (3D) has been confirmed by using the Mie scattering model[18, 19]. However, some properties of the cloak such as the energy transport velocity distributions have not been given explicitly. These velocity distributions are in fact important to study the cloaking effect near the working frequency. Leonhardt has pointed out that time delays are unavoidable for his 2D cloak due to the spatial extension of the refractive-index profile[2]. In this paper, we will consider the time delays and the energy transport velocities in Pendry's 3D invisibility cloaks. We shall limit ourselves to the consideration of cloaking materials that are nearly "ideal" (or very weakly absorbing so that the absorption can be ignored in the discussion). Otherwise the consideration of energy transport velocity is not particularly useful. This is a reasonable assumption for if the cloaking material is strongly absorbing in the operation frequency, it cannot serve as an invisibility cloak in the usual sense of the word.

Pendry's 3D ideal cloaks have the following parameters at the operation frequency,

$$\varepsilon_\theta(\omega_0) = \varepsilon_\varphi(\omega_0) = \mu_\theta(\omega_0) = \mu_\varphi(\omega_0) = \frac{b}{b-a},$$
$$\varepsilon_r(\omega_0) = \mu_r(\omega_0) = \frac{b}{b-a}(\frac{r-a}{r})^2,$$
(1)

where $a$ and $b$ are the inner and outer radii of the 3D cloak[1].

As $b/(b-a)$ is large than 1, we may suppose that $\varepsilon_\theta(\omega)$, $\varepsilon_\varphi(\omega)$, $\mu_\theta(\omega)$ and $\mu_\varphi(\omega)$ are very slow varying functions (treated as constants for simplicity) over a certain frequency range near $\omega_0$, so that

$$\varepsilon_\theta(\omega) = \varepsilon_\varphi(\omega) = \mu_\theta(\omega) = \mu_\varphi(\omega) = \frac{b}{b-a}. \qquad (2)$$

For the radial direction, $\varepsilon_r(\omega_0) = \mu_r(\omega_0) < 1$. Such material properties are typically obtained through some kind of resonance behavior that is excited along the radial direction, and as such, $\varepsilon_r(\omega)$ and $\mu_r(\omega)$ should be treated as explicitly frequency dependent. We consider a general scenario in which they are two different functions of frequencies, $\varepsilon_r(\omega) \neq \mu_r(\omega)$.

In order to provide a visual display of the results, we need to put in some specific functional forms of the permittivity and the permeability. Suppose that $\mu_r(\omega)$ follows the Lorentz model, $\mu_r(\omega) = 1 + \frac{f_m^2}{\omega_m^2 - \omega^2 - i\gamma_m \omega}$. If we omit the absorption, i. e. $\gamma_m = 0$, and we choose for simplicity $\omega_m = \omega_0/2$, from $\mu_r(\omega_0) = \frac{b}{b-a}(\frac{r-a}{r})^2$, we can obtain $f_m$ so that,

$$\mu_r(\omega) = 1 + \frac{3(1 - \frac{b}{b-a}(\frac{r-a}{r})^2)\omega_0^2}{\omega_0^2 - 4\omega^2}.$$

Likewise, the $\varepsilon_r(\omega)$ follows the Drude model, $\varepsilon_r(\omega) = 1 - \frac{f_e^2}{\omega(\omega + i\gamma_e)}$. If we omit the absorption, i. e. $\gamma_e = 0$, from $\varepsilon_r(\omega_0) = \frac{b}{b-a}(\frac{r-a}{r})^2$, we can obtain $f_e$ so that,

$$\varepsilon_r(\omega) = 1 - \frac{(1 - \frac{b}{b-a}(\frac{r-a}{r})^2)\omega_0^2}{\omega^2}.$$

Now we consider the energy transport velocity inside the cloak. Let the wave vector of incident plane wave be along z-direction, and let the electric field be along the x-direction, the

incident field is then

$$\begin{aligned}\vec{E}^{inc} &= \hat{x}\, E_0 \exp(i\frac{\omega}{c} r\cos\theta) = E_0 \exp(i\frac{\omega}{c} r\cos\theta)\sin\theta\cos\varphi\,\hat{r} \\ &+ E_0 \exp(i\frac{\omega}{c} r\cos\theta)\cos\theta\cos\varphi\hat{\theta} - E_0 \exp(i\frac{\omega}{c} r\cos\theta)\sin\varphi\hat{\varphi}.\end{aligned} \quad (3)$$

The electric field inside the cloak can be obtained explicitly as[16],

$$\begin{aligned}\vec{E} &= \frac{b}{b-a} E_0 \exp(i\frac{\omega}{c}\frac{b}{b-a}(r-a)\cos\theta)\sin\theta\cos\varphi\hat{r} \\ &+ \frac{b}{b-a}\frac{r-a}{r} E_0 \exp(i\frac{\omega}{c}\frac{b}{b-a}(r-a)\cos\theta)\cos\theta\cos\varphi\hat{\theta} \\ &- \frac{b}{b-a}\frac{r-a}{r} E_0 \exp(i\frac{\omega}{c}\frac{b}{b-a}(r-a)\cos\theta)\sin\varphi\hat{\varphi}.\end{aligned} \quad (4)$$

Likewise, we can obtain the magnetic field $\vec{H}$ and the Poynting vector $\vec{S}$ inside the cloak. From the definition of the energy transport velocity $v_e = |\vec{S}|/U$, we can obtain the energy transport velocity. Here the energy density $U$ is[20],

$$U = \frac{1}{4}\mathrm{Re}[(\varepsilon_0 \vec{E}\bullet \begin{bmatrix} \frac{\partial(\omega\varepsilon_r)}{\partial\omega} & 0 & 0 \\ 0 & \varepsilon_\theta & 0 \\ 0 & 0 & \varepsilon_\varphi \end{bmatrix}\vec{E}^* + \mu_0\vec{H}\bullet \begin{bmatrix} \frac{\partial(\omega\mu_r)}{\partial\omega} & 0 & 0 \\ 0 & \mu_\theta & 0 \\ 0 & 0 & \mu_\varphi \end{bmatrix}\vec{H}^*)], \quad (5)$$

the above tensors are both expressed in the spherical coordinate system. The energy transport velocity is then,

$$\frac{v_e}{c} = \frac{2[((\frac{b}{b-a})^2(\frac{r-a}{r})^2\cos\theta)^2 + ((\frac{b}{b-a})^2(\frac{r-a}{r})\sin\theta)^2]^{1/2}}{(\frac{b}{b-a}\sin\theta)^2[(\varepsilon_r + \omega\frac{d\varepsilon_r}{d\omega})(\cos\varphi)^2 + (\mu_r + \omega\frac{d\mu_r}{d\omega})(\sin\varphi)^2]} \qquad (6)$$
$$+\frac{b}{b-a}(\frac{b}{b-a}\frac{r-a}{r}\cos\theta)^2 + \frac{b}{b-a}(\frac{b}{b-a}\frac{r-a}{r})^2$$

Fig. 1(a-e) shows the energy velocity distributions for different planes, ($\varphi = 0^o$, $30^o$, $45^o$, $60^o$ and $90^o$) of a fixed polarization (E field along x-direction). In the figure, we set $a = 27.1mm$ and $b = 58.9mm$, $\omega_0 = 2\pi \times 8.5 GHz$, so that the parameters are relevant for the experiment described in Ref [15]. Other choices of parameters of course will not change the physics. We see that the velocity distribution have rather similar distributions. So, we need only discuss the results in fig. 1(c). We see from fig. 1(c) that the energy transport velocity is always smaller than c (velocity of light in vacuum). The energy transport velocity is larger near $\theta = 0$ or $\pi$ (along the incident z direction). We note in particular that the energy transport velocity is much smaller (approaching zero) near the inner boundary, which will cause that rays near the inner boundary to reach a target plane much more slowly than that are far away from the inner boundary. We will see in later sections that a Hamilton's optics consideration reach the same conclusion. The energy transport velocity actually has a rather peculiar distribution near the inner boundary. We found that at the inner boundary (when $r = a$ and $\theta \neq 0$ or $\pi$), $v_e / c = 0$; but the velocity is finite at two points ($v_e / c = (b-a)/b$ when $\theta = 0$ or $\pi$). When light impinges near the inner boundary, its propagation direction will have a change of angle and the ray will ride along the spherical surface of the inner boundary with an energy transport velocity that approaches zero. A ray would take nearly infinite time to reach from one side of the cloak to another side if it crosses the origin of cloak.

To corroborate with the above energy velocity distribution consideration, we now come to the Hamiltonian optics description.

For the target frequency $\omega_0$, the dispersion relation is,

$$\frac{\kappa_1^2}{(\frac{b}{b-a})^2} + \frac{\kappa_2^2 + \kappa_3^2}{(\frac{b}{b-a})^2(\frac{r-a}{r})^2} - 1 = 0, \tag{7}$$

where $\vec{\kappa} = \kappa_1 \hat{r} + \kappa_2 \hat{\theta} + \kappa_3 \hat{\varphi}$, $\vec{k} = \frac{\omega}{c}\vec{\kappa}$ is the real wave vector, and $\hat{r}$, $\hat{\theta}$ and $\hat{\varphi}$ are unit vectors in the spherical coordinate systems for every point in real space inside the cloak.

The Hamiltonian of the 3D ideal cloak[21] is,

$$H = \frac{1}{2}\vec{\kappa}\bullet\vec{\kappa} - \frac{1}{2}\frac{2ar-a^2}{r^4}(\vec{r}\bullet\vec{\kappa})^2 - \frac{1}{2}\left[\frac{b(r-a)}{r(b-a)}\right]^2. \tag{8}$$

The equations of motion are then,

$$\frac{d\vec{r}}{d\tau} = \frac{\partial H}{\partial \vec{\kappa}}, \quad \frac{d\vec{\kappa}}{d\tau} = -\frac{\partial H}{\partial \vec{r}}, \tag{9}$$

where $\vec{r} = r\hat{r}$, and $\tau$ parameterizes the paths, and the path parameter $\tau$ is related to the propagation time $t$ by the following relationship $cdt = f(\vec{r})d\tau$ so that,

$$\frac{d\vec{r}}{f(\vec{r})d\tau} = \frac{d\vec{r}}{cdt} = \frac{\vec{v}_g}{c} = \frac{\frac{\partial H}{\partial \vec{k}}}{-\frac{\partial H}{\partial \omega}}. \tag{10}$$

We can obtain six coupled first order ordinary differential equations from the equations of motion in the spherical coordinate system,

$$\kappa_1 (\frac{r-a}{r})^2 = \dot{r}\frac{f(\bar{r})}{c},$$

$$\kappa_2 = r\dot{\theta}\frac{f(\bar{r})}{c},$$

$$\kappa_3 = r\sin\theta\dot{\varphi}\frac{f(\bar{r})}{c},$$

$$(\frac{a^2-ar}{r^3})(\kappa_1^2 - (\frac{b}{b-a})^2) = \dot{\kappa}_1\frac{f(\bar{r})}{c} - \kappa_2\dot{\theta}\frac{f(\bar{r})}{c} - \kappa_3\dot{\varphi}\frac{f(\bar{r})}{c}\sin\theta, \quad (11)$$

$$\frac{2ar-a^2}{r^3}\kappa_1\kappa_2 = \dot{\kappa}_2\frac{f(\bar{r})}{c} + \kappa_1\dot{\theta}\frac{f(\bar{r})}{c} - \kappa_3\dot{\varphi}\frac{f(\bar{r})}{c}\cos\theta,$$

$$\frac{2ar-a^2}{r^3}\kappa_1\kappa_3 = \dot{\kappa}_3\frac{f(\bar{r})}{c} + \kappa_1\dot{\varphi}\frac{f(\bar{r})}{c}\sin\theta + \kappa_2\dot{\varphi}\frac{f(\bar{r})}{c}\cos\theta,$$

where $\dot{h} = dh/dt$, $h = r$, $\theta$, $\varphi$, $\kappa_1$, $\kappa_2$ and $\kappa_3$. The above equations can be integrated numerically using the Runge-Kutta method if the initial conditions are known.

From Eq. (11) we can obtain the relationship between the k-vector and the position coordinates (see Appendix A),

$$\kappa_1 = \frac{b}{b-a}\cos\theta, \quad \kappa_2 = -\frac{b}{b-a}\frac{r-a}{r}\sin\theta, \quad \kappa_3 = 0. \quad (12)$$

Using the methods in the Ref [17], one can obtain the defined group velocity from the dispersion relationship. Taking the above given specific functional forms of the permittivity and the permeability and with the knowledge of k-vector (Eq. 12), we have,

$$\frac{\bar{v}_g}{c} = \frac{(\frac{r-a}{r})^2\cos\theta\hat{r} - \frac{r-a}{r}\sin\theta\hat{\theta}}{\frac{b}{b-a}(\frac{r-a}{r})^2 + \frac{7}{6}(1-\frac{b}{b-a}(\frac{r-a}{r})^2)\sin^2\theta}. \quad (13)$$

Then from Eq. (10) and Eq. (11), we have

$$f\frac{\vec{v}_g}{c} = \frac{d\vec{r}}{d\tau} = \frac{dr}{d\tau}\hat{r} + r\frac{d\theta}{d\tau}\hat{\theta} = \kappa_1(\frac{r-a}{r})^2\hat{r} + \kappa_2\hat{\theta}$$
$$= \frac{b}{b-a}((\frac{r-a}{r})^2\cos\theta\hat{r} - \frac{r-a}{r}\sin\theta\hat{\theta}). \tag{14}$$

Compare Eq. (13) and Eq. (14), we can obtain that,

$$f(\vec{r}) = (\frac{b}{b-a})^2(\frac{r-a}{r})^2 + \frac{7}{6}\frac{b}{b-a}(1 - \frac{b}{b-a}(\frac{r-a}{r})^2)\sin^2\theta \tag{15}$$

We now have all the information needed to calculate the time needed to cover a certain optical path within the cloak for the target frequency. Let there be a beam that is incident from the $-z$ direction. When the beam crosses a point on the outer boundary of the cloak at $r=b$, where $x_0 = b\sin\theta_0\cos\varphi_0$, $y_0 = b\sin\theta_0\sin\varphi_0$, $z_0 = b\cos\theta_0$, the parameter is $t=t_0$, so that $r|_{t=t_0}=b$, $\theta|_{t=t_0}=\theta_0$, $\varphi|_{t=t_0}=\varphi_0$. Consider the boundary condition, $(\vec{\kappa}_0 - \vec{\kappa})\times\hat{r}=0$ with $\vec{\kappa}_0 = \hat{z}$ in the free space, then we obtain, $\kappa_2|_{t=t_0}=-\sin\theta_0$, $\kappa_3|_{t=t_0}=0$, together with the dispersion relationship, one can obtain the detail information of the path of the ray in the cloak.

As a concrete example, we suppose that $a=1\ units$, $b=2\ units$, and $x|_{t=0}=x_0$, $y|_{t=0}=0$, $z|_{t=0}=-3\ units$ where we choose a set of $x_0$ from $-1.75\ units$ to $1.75\ units$, the ray is traced from $t=0$ to $t=12/c$. From fig. 2, we shall see that depending on their incident positions, different rays have propagated to different positions in space. Specifically, we see that when $t=12/c$, six rays ($x_0 = \pm 1.75\ units$, $\pm 1.5\ units$ and $\pm 1.25\ units$) have already passed through the plane $z=3\ units$, while other rays that are closer to the inner boundary are still "on their way" to reach the plane. It is visually obvious in Fig. 2 that the rays which are far away from the inner boundary travels faster so that they will reach a plane outside the cloak earlier while the rays which are closer to the inner boundary will spend much more time inside the cloak

and they will reach the target plane at a much later time. The closer to the inner boundary, the slower is the speed of the beam. These phenomena are consistent with the peculiar energy velocity distributions shown above.

Furthermore, we can obtain the relationship between the time spent from plane $z = -3\ units$ to $z = 3\ units$ and $x_0$ (see Appendix A),

$$c\Delta t = 6\ units + [\frac{7a^2}{6(b-a)b} + \frac{14a}{3b}]\sqrt{b^2 - x_0^2} + \frac{7\pi a^2 b}{12(b-a)x_0} \\ - \frac{7\pi a x_0}{6b} - \frac{7a^2 b}{6(b-a)x_0}\arcsin\frac{x_0}{b} + \frac{7ax_0}{3b}\arcsin\frac{x_0}{b} \qquad (16)$$

We note the limit $\lim_{x_0 \to 0} c\Delta t(x_0) = +\infty$, which means when $x_0$ goes to zero, $\Delta t$ will tend to infinity. Therefore for a ray that cuts across the origin of the cloak, it will never reach the target plane within a geometric optics description, or alternatively, the ideal cloak effect should take a very long time to set up dynamically. In a finite interval of time, the cloak will be still scatter[22]. The Hamiltonian optics results are consistent with the energy velocity considerations.

In summary, we have given explicit expressions for the energy transport velocity distribution inside a 3D ideal cloak, and we computed the velocity for some specific forms of permittivity and permeability. We found that the energy transport velocity is always smaller than the velocity of light in vacuum and has larger values along the incident direction in the line joining the origin of the cloak. In addition, the energy transport velocity near the inner boundary is much smaller or becomes zero at the inner boundary, and rays near the inner boundary will take a much longer time to reach the other side of cloak. For a beam cutting cross the origin of the cloak will take infinite time to pass through the cloak from the view of geometric optics.

We thank Prof. Xunya Jiang and Dr. Wei Ren for many helpful discussions. HYC also thanks Prof. Ulf Leohardt for many helpful comments. This work was supported by Hong Kong Central Allocation grant HKUST3/06C. Computation resources are supported by Shun Hing Education and Charity Fund.

**Figure Captions:**

Fig. 1. The energy velocity distribution near the cloak for different plane. (a) $\varphi = 0^o$; (b) $\varphi = 30^o$; (c) $\varphi = 45^o$; (d) $\varphi = 60^o$; (e) $\varphi = 90^o$.

Fig. 2. (Color online) The ray tracing for different beam, $x_0 = \pm 1.75$ *units*, $\pm 1.5$ *units*, $\pm 1.25$ *units*, $\pm 1$ *units*, $\pm 0.75$ *units*, $\pm 0.5$ *units*, $\pm 0.25$ *units* and $\pm 0.1$ *units*, $t$ traced from 0 to $12/c$.

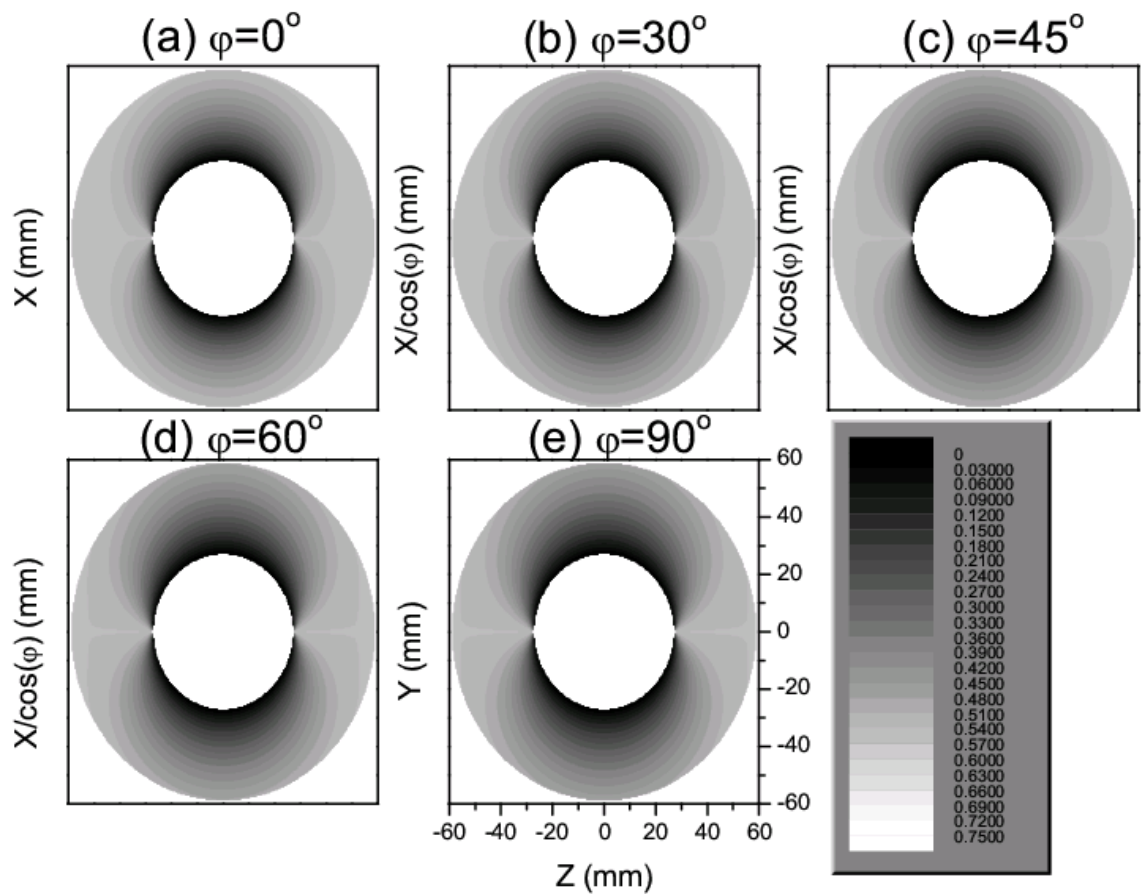

Fig. 1.

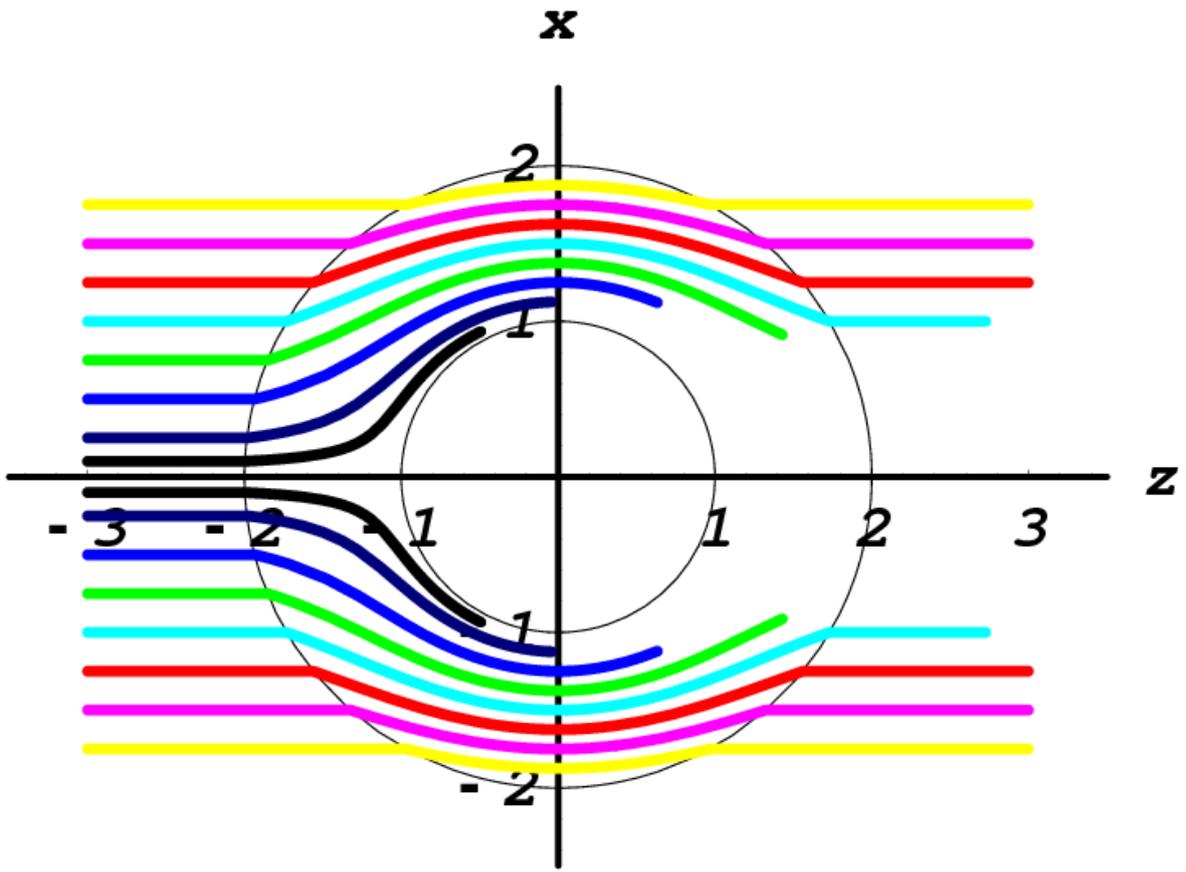

Fig. 2.

# Appendix A: The detailed derivation of Eq. (12) and Eq. (16)

Suppose that the path of the ray in free space is parameterized by,

$$r\sin\theta\cos\varphi = x_0, \quad r\sin\theta\sin\varphi = y_0. \tag{A. 1}$$

According to the principle of the transformation media "cloak", the ray in the cloak follows the trajectory,

$$\frac{b}{b-a}(r-a)\sin\theta\cos\varphi = x_0, \quad \frac{b}{b-a}(r-a)\sin\theta\sin\varphi = y_0, \tag{A. 2}$$

i.e. $\theta$ and $\varphi$ are kept unchanged, while $r$ are mapped into $\frac{b}{b-a}(r-a)$, the above is the starting point of creating a cloak using the transformation media method.

Take the total differential form of Eq. (A. 2), we have,

$$\sin\theta\cos\varphi dr + (r-a)\cos\theta\cos\varphi d\theta - (r-a)\sin\theta\sin\varphi d\varphi = 0,$$
$$\sin\theta\sin\varphi dr + (r-a)\cos\theta\sin\varphi d\theta + (r-a)\sin\theta\cos\varphi d\varphi = 0, \tag{A. 3}$$

which are,

$$\sin\theta\cos\varphi + (r-a)\cos\theta\cos\varphi\frac{d\theta}{dr} - (r-a)\sin\theta\sin\varphi\frac{d\varphi}{dr} = 0,$$
$$\sin\theta\sin\varphi + (r-a)\cos\theta\sin\varphi\frac{d\theta}{dr} + (r-a)\sin\theta\cos\varphi\frac{d\varphi}{dr} = 0, \tag{A. 4}$$

therefore we have,

$$\sin\theta + (r-a)\cos\theta \frac{d\theta}{dr} = 0,$$

$$(r-a)\sin\theta \frac{d\varphi}{dr} = 0,$$

(A. 5)

or equivalently,

$$\frac{d\theta}{dr} = -\frac{1}{r-a}\frac{\sin\theta}{\cos\theta},$$

$$\frac{d\varphi}{dr} = 0.$$

(A. 6)

From the first three equations in Eq. (11),

$$\frac{d\theta}{dr} = \frac{\dot\theta}{\dot r} = \frac{\kappa_2}{\kappa_1}(\frac{r}{r-a})^2\frac{1}{r},$$

$$\frac{d\varphi}{dr} = \frac{\dot\varphi}{\dot r} = \frac{\kappa_3}{\kappa_1}(\frac{r}{r-a})^2\frac{1}{r\sin\theta}.$$

(A. 7)

So the following are obtained,

$$\frac{\kappa_2}{\kappa_1} = -\frac{\sin\theta}{\cos\theta}\frac{r-a}{r},$$

$$\kappa_3 = 0.$$

(A. 8)

Then from the dispersion relationship Eq. (7), we have,

$$\kappa_1 = \frac{b}{b-a}\cos\theta,\ \kappa_2 = -\frac{b}{b-a}\frac{r-a}{r}\sin\theta,\ \kappa_3 = 0.$$

(A. 9)

This equation describes the dimensionless k-vector everywhere in the cloak (Eq. (12)).

From $\kappa_3 = 0$ or $\dot\varphi = 0$, we know that every coordinate $\varphi$ in a cloak is equal to a constant

$\varphi_0$ which is the coordinate of a point in the outer boundary of the cloak where the incident ray first impinges. Then we would rewrite Eq. (A.2) into,

$$(r-a)\sin\theta = \frac{b-a}{b}\frac{x_0}{\cos\varphi_0} = \frac{b-a}{b}\frac{y_0}{\sin\varphi_0} = d, \tag{A.10}$$

which is,

$$r = a + \frac{d}{\sin\theta}, \tag{A.11}$$

From $\kappa_2 = r\dot{\theta}\frac{f(\bar{r})}{c} = -\frac{b}{b-a}\frac{r-a}{r}\sin\theta$, we obtain,

$$\frac{d\theta}{cdt} = \frac{-\frac{b}{b-a}\frac{r-a}{r^2}\sin\theta}{f(\bar{r})}, \tag{A.12}$$

or equivalently,

$$cdt = -d\theta\,[\frac{bd}{b-a}\frac{1}{\sin^2\theta} + \frac{7}{6}[\frac{(a\sin\theta+d)^2}{d} - \frac{bd}{b-a}]]. \tag{A.13}$$

After integration, solving the above equation gives,

$$\begin{aligned}c(t_0 - t) =& \frac{bd}{b-a}[\cot\theta_0 - \cot\theta] - \frac{7}{24}\frac{a^2}{d}(\sin 2\theta - \sin 2\theta_0)\\&+ \frac{7}{3}a[\cos\theta_0 - \cos\theta] + [\frac{7}{12}\frac{a^2}{d} - \frac{7}{6}\frac{ad}{b-a}](\theta - \theta_0).\end{aligned} \tag{A.14}$$

Furthermore, we obtain the relationship between the time spent from plane $z = -3$ *units* to $z = 3$ *units* and $x_0$ (Eq. (16)).